\documentclass[aps,prl,twocolumn,longbibliography,showpacs,amsmath,amssymb,10pt]{revtex4-1} 
\usepackage{graphicx,subfig,xcolor}
\usepackage{array}
\usepackage{ragged2e}
\usepackage{cancel}
\definecolor{darkblue}{rgb}{0.1,0.1,.7}
\usepackage[colorlinks, linkcolor=darkblue, citecolor=darkblue, urlcolor=darkblue, linktocpage]{hyperref} 
\newcolumntype{L}[1]{>{\raggedright\let\newline\\\arraybackslash\hspace{0pt}}m{#1}}
\newcolumntype{C}[1]{>{\centering\let\newline\\\arraybackslash\hspace{0pt}}m{#1}}
\newcolumntype{R}[1]{>{\raggedleft\let\newline\\\arraybackslash\hspace{0pt}}m{#1}}
\usepackage{manfnt,pifont,dsfont}
\usepackage{ulem}

\newcommand\del{\partial}

\newcommand\cO{{\cal O}}

\newcommand{\be}{\begin{equation}}
\newcommand{\ee}{\end{equation}}
\newcommand{\bea}{\begin{eqnarray}}
\newcommand{\eea}{\end{eqnarray}}
\newcommand\restr[2]{{\left.\kern-\nulldelimiterspace#1\vphantom{\big|}\right|_{#2}}}

\newcommand{\reef}[1]{(\ref{#1})}

\newcommand{\beq}{\begin{equation}} 
\newcommand{\eeq}{\end{equation}}
\def\del {\partial} 
 
\def\bZ {\mathbb{Z}} 
\def\bR {\mathbb{R}}

\def\calH {{\cal H}} 
 
\def\calE {{\cal E}} 
\def\bZ {\mathbb{Z}} 
 
\def\bZ {\mathbb{Z}} 
 
\def\bP {\mathbb{P}} 
\def\half{{\textstyle\frac 12}}

\def\ge{\geqslant}
\def\le{\leqslant}
\def\geq{\geqslant}
\def\leq{\leqslant}

\def \del{\partial}

\newcommand{\NO}[1]{{:\!#1\!:}}

\def\phys{{\rm ph}}

\newcommand{\braket}[3]{\langle #1|#2|#3 \rangle}

\newcommand{\ket}[1]{|#1\rangle}
\newcommand{\bra}[1]{\langle #1|}

 \def\cE{{\cal E}}

\begin{document}
\vspace*{-1in} 
\begin{flushright}
CERN-TH-2017-132
\end{flushright}
\title{High-Precision Calculations in Strongly Coupled Quantum Field Theory\\
with Next-to-Leading-Order Renormalized Hamiltonian Truncation}
\author{Joan Elias-Mir\'o$^a$, Slava Rychkov$^{b,c}$,  Lorenzo G. Vitale$^{d,e}$ } 
\affiliation{
$^{a}$ SISSA/ISAS and INFN, I-34136 Trieste, Italy\\
$^b$  CERN, Theoretical Physics Department, 1211 Geneva 23, Switzerland\\
$^c$  Laboratoire de Physique Th\'eorique de l'\'Ecole Normale Sup\'erieure,  \\
PSL Research University,  CNRS,  Sorbonne Universit\'es, UPMC
Univ.\,Paris 06,  \\
24 rue Lhomond, 75231 Paris Cedex 05, France
\\
$^d$ Institut de Th\'eorie des Ph\'enom\`enes Physiques, EPFL, CH-1015 Lausanne, Switzerland\\
$^e$ Department of Physics, Boston University, Boston, MA 02215
}

\begin{abstract}
Hamiltonian Truncation (a.k.a.~Truncated Spectrum Approach) is an efficient numerical technique to solve strongly coupled QFTs in $d=2$ spacetime dimensions. 
Further theoretical developments are needed to increase its accuracy and the range of applicability.
With this goal in mind, here we present a new variant of Hamiltonian Truncation which exhibits smaller dependence on the UV cutoff than other existing implementations, and yields more accurate spectra. 
The key idea for achieving this consists in integrating
out exactly a certain class of high energy states, which corresponds to performing
renormalization at the cubic order in the interaction strength.
We test the new method on the strongly coupled two-dimensional quartic scalar theory. Our work will also be useful for the future goal of extending Hamiltonian Truncation to higher dimensions $d\ge 3$.
\end{abstract}
\maketitle
\pagebreak
\nopagebreak
{\bf Introduction.} Many interesting strongly interacting 
Quantum Field Theories (QFTs) are not amenable to analytical treatment. 
Such theories are often studied via Lattice Monte Carlo 
(LMC) numerical simulations, starting from the discretized Euclidean action. 
However, LMC has some drawbacks, for example it cannot easily compute real-time observables, 
it is rather computationally expensive, and it cannot directly 
describe renormalization group (RG) flows starting from interacting fixed points.
Therefore, it is worth exploring other numerical approaches to 
strongly interacting QFTs. One promising alternative is provided by the Hamiltonian methods, which look for the eigenstates of the quantum Hamiltonian. These methods use various finite-dimensional approximations to the full infinite-dimensional QFT Hilbert space. Notable examples are the methods using Matrix Product States \cite{White:1993zza, MPSreps}
and more general Tensor Networks \cite{ShiDuanVidal} such as MERA \cite{MERA} or PEPS \cite{PEPS}.
In this paper we will be concerned with another representative of this group of methods---Hamiltonian Truncation (HT), also known as the Truncated Spectrum (or Space) Approach, which is a direct generalization of the variational Rayleigh-Ritz (RR) method from quantum mechanics. This method goes back to the seminal work of Yurov and Al.~Zamolodchikov \cite{Yurov:1989yu,Yurov:1991my} and has since been applied in many contexts.
See \cite{Konik-review} for a recent extensive review and the bibliography. 

The idea of HT is simple. 
The QFT Hamiltonian operator $H$ is split as $H_0+V$ where $H_0$ is an exactly solvable Hamiltonian whose eigenstates form the basis of the Hilbert space. One quantizes at surfaces of constant time and works in finite volume so that the spectrum is discrete 
\footnote{In relativistic QFTs one can also quantize on surfaces of constant light-cone coordinate. This \textit{light front quantization} \cite{Brodsky:1997de} is also used in numerical solutions of strongly coupled QFTs via a version of HT; some recent work is \cite{Katz:2013qua,Katz:2014uoa,Chabysheva:2015ynr,Burkardt:2016ffk,Katz:2016hxp,Anand:2017yij}. The structure of the unperturbed Hilbert space is different from the equal time case, which leads to important differences in the numerical procedure. All technical claims in this work will refer exclusively to the equal time quantization.}. 
The Hilbert space is then truncated to the low-lying eigenvectors of $H_0$.
The matrix of $H$ in this truncated Hilbert space is diagonalized exactly on a computer, 
to find the low-energy spectrum of interacting eigenstates.

As was understood early on \cite{Klassen:1991ze}, the numerical convergence of the HT depends crucially on 
the scaling dimension $\Delta_V$ of the interaction $V$. If the interaction is strongly relevant, in the RG sense,
then HT converges fast, but convergence rate worsens as $\Delta_V$ increases.  This is a limitation of the method. For interaction dimensions larger that $d/2$, naive HT actually diverges \cite{Klassen:1991ze}. To ensure the convergence, we will assume here that 
\beq
\Delta_V<d/2\,. \label{assumption}
\eeq
\vspace{-0.5cm}

Another limitation of HT, as of many variational methods in general,
is that the Hilbert space grows exponentially with the cutoff. Specifically, the dimension grows as $\exp({C E_T^\alpha})$, where $C>0$ is a theory-dependent constant and $E_T$ is the energy cutoff on the $H_0$ spectrum. The exponent typically depends on the spacetime dimension as $\alpha=1-1/d$, so this problem becomes more severe in higher $d$. These two limitations are the main reason why the HT has been so far applied mainly in $d=2$.

Motivated by the need to mitigate the above limitations, the recent works \cite{Hogervorst:2014rta,Lorenzo1,Lorenzo2,Elias-Miro:2015bqk} (following notably \cite{Giokas:2011ix}; see also \cite{Feverati:2006ni,Watts:2011cr, Lencses:2014tba}) started 
developing the theory of renormalized HT, in which high-energy
modes are not simply truncated away, but integrated out to produce an effective low energy Hamiltonian.
As a result the convergence is improved. Renormalized HT has been applied in several strongly coupled QFT studies in $d=2$ \cite{Giokas:2011ix,Lencses:2014tba,Lorenzo1,Lorenzo2,Elias-Miro:2015bqk} and in one study in $d=2.5$ \cite{Hogervorst:2014rta}.
  We hope that in the future Hamiltonian Truncation will develop into an accurate numerical method, applicable also in $d \ge 3$. Here we will take another step towards this goal by proposing a novel 
and still more accurate approach to renormalization. A more detailed technical account of our work will appear elsewhere \cite{JSL2}. 

{\bf Setup.}  
Consider the Hamiltonian $H$ of a QFT in finite volume,   which we assume can be split 
into a solvable part
$H_0$, whose eigenfunction and eigenvectors are known, plus an interaction $V$, whose matrix elements  are computable in the basis of eigenstates of $H_0$:
\begin{equation}
H = H_0 + V \,, \quad H_0 \ket{i} = E_i \ket{i} \,, 
\quad V_{ij} = \braket{i}{V}{j}\,.
\label{eq:ham}
\end{equation}
The $H_0$ can represent a free or an integrable Hamiltonian, or 
the Hamiltonian of a conformal field theory (CFT) on the cylinder $S^{d-1}\times \bR$. 
We assume that its finite volume spectrum is discrete, which is the case for most $H_0$'s of interest. 
 Notice that although numerical HT calculations are performed in finite volume, infinite volume observables can then be extracted via controlled extrapolation.

The spectrum of the interacting theory is found by solving the eigenvalue equation:
\beq
H.c=\calE  c,\quad c\in \calH\, ,
\label{p1}
\eeq
where $\calE$ is the energy of a given state, and $c$ the corresponding eigenvector living in the Hilbert space $\calH$ spanned by the eigenstates of $H_0$. Eq.~\eqref{p1} is infinite-dimensional and cannot be solved on a computer.
So we split $\calH$ into a finite-dimensional
``low-energy'' part $\calH_l$ and a ``high-energy'' part $\calH_h$. Motivated by effective field theory, a natural choice is to include into $\calH_l$ all the states with 
energy below a given cutoff $E_i \le E_T$,  which plays the role of a UV cutoff. This should provide a good approximation for the interacting eigenstates 
with energy well below the cutoff. Different types of cutoff are possible but will not be considered here.
We then project the eigenvalue equation onto those subspaces:
\begin{align}
H_{ll}.c_l + V_{lh}.c_h&=\calE c_l\,, \label{loeig}\\
V_{hl}.c_l + H_{hh}.c_h&=\calE c_h \label{hieig}\, ,
\end{align}
 where $c=(c_l ,c_h)$ is the low/high energy split of $c$, i.e. 
$c_l = P_l c$, $c_h = P_h c $, where $P_l$, $P_h$ are the projectors on 
$\calH_l$, $\calH_h$. Similarly, $H_{ll} = P_l H P_l$, and so on.

The {\it raw} HT consists in throwing out all the states 
in $\calH_h$ and solving the eigenvalue equation,
\begin{equation}
H_{ll}.c_l =\calE_{\rm raw} c_l\, . \label{raweig}\\
\end{equation}
By the min-max theorem, as the cutoff $E_T$ is increased, the   eigenvalues
$\calE_{\rm raw}$ approach the exact eigenvalues $\calE$ from above.
As shown in \cite{Hogervorst:2014rta}, the raw HT numerical spectrum is expected to converge with polynomial rate $1/E_T^{\rho}$, with $\rho = d-2\Delta_V>0$ by our assumption \reef{assumption}.  
This polynomial convergence must compete with the exponential growth of states in the Hilbert space.

It is possible to do better than in \reef{raweig}. Instead of simply truncating 
\reef{loeig}, we use \reef{hieig} to express the high-energy part $c_h$ of the eigenvector in terms of the low-energy part $c_l$:
\be
c_h = (\cE-H_{hh})^{-1}.V_{hl}.c_l \, . \label{optimaltails}
\ee
Plugging this back in \reef{loeig} gives the equation
\be
H_{\rm eff}.c_l = \cE c_l \, , \label{heff}
\ee
where $H_{\rm eff}$ is the effective Hamiltonian operator acting on ${\cal H}_l$. It is given by
\bea
H_{\rm eff} &=& H_{ll}+ \Delta H(\cE) \,,  \label{heff1}
\\ \Delta H (\cE) &=& V_{lh}. (\cE-H_{hh})^{-1}.V_{hl} \, .  \label{DeltaH}
\eea
The solutions of \reef{heff} are equivalent to the solutions of the original eigenvalue problem \reef{p1}.

Eqs.~(\ref{heff1},\ref{DeltaH}) are the starting point of \textit{renormalized} HT 
\footnote{This has to be distinguished from the Numerical Renormalization Group (NRG) improvement of the HT \cite{Konik:2007cb} \`a la Wilson's NRG \cite{Wilson}.  
This method raises the cutoff by adding new chunks of the Hilbert space and tossing away the states which have low overlaps with the interacting eigenstates. Other ideas to extend the reach of HT include \textit{sweeping} and \textit{reordering} (see \cite{Konik-review}). We have not used any of these interesting tricks in our work.}. 
Integrating out the high-energy part of $c_h$ we correct or, as we say, renormalize $H_{ll}$ by $\Delta H$.  
While in general $\Delta H$ cannot be computed exactly, the goal
is to approximate it sufficiently well so that solutions of \reef{heff} become close to the exact eigenenergies. 
The hope is that this can be done keeping the cutoff $E_T$, and therefore the dimension of ${\cal H}_l$, 
relatively low and manageable on a computer.

One natural way to approximate $\Delta H (\cE)$ would be via an expansion
in powers of $V$:
\begin{gather}
\Delta H(\cE)=\sum_{n=2}^\infty  \Delta H_n(\cE)  \label{series1} \, ,\\
\label{seriesdh}
\Delta H_n = V_{lh}\frac{1}{\cE-H_{0hh}}\left(V_{hh}\frac{1}{\cE-H_{0hh}}\right)^{n-2} V_{hl}  \, ,
\end{gather}
truncating it to a fixed order. This is what was done in the previous
works \cite{Hogervorst:2014rta,Lorenzo1,Lorenzo2}, 
where \reef{series1} was truncated to the leading order (LO) $n=2$, and $\Delta H_2$ was computed in an 
analytic {\it local} approximation 
\beq
\Delta H \approx \Delta H_2^\text{local}\,, \label{H2local}
\eeq
 which will be briefly reviewed after Eq.~\reef{genloc} below. 
This was shown to improve significantly the numerical
convergence of the spectrum in $E_T$.
However, in Ref.~\cite{Elias-Miro:2015bqk} it was shown that, first of all, this method is not easily generalizable 
to higher orders and second, increasing the accuracy of the approximation
of $\Delta H_2$ alone does not necessarily improve the convergence.  Furthermore, the naive expansion 
\eqref{seriesdh} is not convergent and there will appear unbounded matrix elements as the power of $V$ is 
increased \footnote{That can be intuitively understood as follows. For each $n \ge 2$, there will be states below the cutoff 
for which the matrix elements of $\Delta H_n$ grow as $\sim (c N)^n E_T$ in absolute value, where $N$ is the occupation number 
of the state and $c$ is a constant. For $N$ big enough, the expansion is therefore not convergent, as the truncated matrix 
element will outgrow the leading order contribution of $H_0$ growing as $\sim E_T$. For a detailed discussion of this point 
see \cite{JSL2}, appendix B.}.

We will now introduce the main novelty of the present paper---an approach to renormalize the truncated Hamiltonian that neatly avoids the problems pointed out by \cite{Elias-Miro:2015bqk}, and leads to a more accurate spectrum than any previous approach. 

{\bf NLO-HT as integrating out tails.} 
Let us rethink Eqs.~(\ref{raweig}-\ref{DeltaH}). Eq.~\reef{raweig} can be viewed as an instance of the 
RR approach, where the full Hamiltonian has been projected on the finite-dimensional
subspace $\calH_l \subset \calH_l \oplus \calH_h$. 
The high-energy Hilbert space $\calH_h$ is infinite-dimensional, but Eq.~\reef{optimaltails} implies that we don't need all of it. Indeed, this equation says that one could retrieve the exact result  
by truncating $\calH_h$ to a finite dimensional subspace 
$\calH_t$ spanned by the vectors
\be 
  (\cE-H_{hh})^{-1}.V_{hl} \ket{i} \,, \qquad \ket{i} \in \calH_l  \, . 
 \label{opwf}
\ee
Of course, these states are impossible to compute exactly, so let us approximate them by setting $H_{hh}\approx H_{0hh}$, i.e. 
\be
\ket{\Psi_i }\equiv (\cE_*-H_0)^{-1}.V_{hl} \ket{i} \, , \label{realtails}
\ee
with $\cE_*$ a parameter that will be set close to $\cE$.
We call these $\ket{\Psi_i}$'s \textit{tail states}, as their linear combination approximate 
the high-energy ``tail'' of the eigenvectors.
We next consider the eigenvalue equation for the Hamiltonian \reef{p1} projected on the space spanned by $\{\ket{i},\,\ket{\Psi_i}\}$:
\bea
H_{ll}.c_l + H_{lt}.c_t &=& \cE_{\rm RR} c_l \, ,  \label{sys1} \\
H_{tl}.c_l+H_{tt}.c_t &=& \cE_{\rm RR} G. c_t \label{sys2}\, ,
\eea
where $G_{ij}=\langle\Psi_i|\Psi_j\rangle$ is the Gram matrix of the tail states, which are not orthonormal, and
\bea
(H_{lt})_{ij} &=& \bra{i}H\ket{\Psi_j}= [\Delta H_2]_{ij} \, ,  \label{m1}\\
(H_{tt})_{ij}&=&\bra{\Psi_i}H\ket{\Psi_j} =[\Delta H_3-\Delta H_2+\cE_* G]_{ij} \, .\quad \  \
\label{m2}  
\eea
Here $\Delta H_2$ and $\Delta H_3$ are the same as above with $\calE\to\calE_*$.

Assuming that the operators (\ref{m1},\ref{m2}) can be evaluated to high accuracy, 
one can diagonalize
(\ref{sys1},\ref{sys2}) numerically on a computer and obtain the Rayleigh-Ritz eigenvalues $\cE_\text{RR}$. By construction, these eigenvalues have variational interpretation with an ansatz enlarged with respect to the raw HT, implying via the min-max theorem that 
$\cE \le \cE_\text{RR} \le \cE_\text{raw}$ 
\footnote{In this work, we introduce a tail state for each \unexpanded{$\ket{i}\in\calH_l$}. This limits the number of states we include
in the basis, as the full matrix ($\Delta H_2 - \Delta H_3$) needs to be inverted
over the space of tail states. On the contrary, the low-energy diagonalization of 
$H_{l l} + \Delta \tilde{H}$ is performed efficiently via the Lanczos method.
This suggests that more efficient numerics could be achieved by reducing 
the number of tails states \unexpanded{$\ket{\Psi_i}$}; see \cite{JSL2} for a discussion.}. 

Let us transform equations (\ref{sys1},\ref{sys2}) further 
by integrating out the tail states. 
Substituting $c_t$ from \reef{sys2} into \reef{sys1} we get an 
equivalent equation for the RR spectrum:
\be
[H_{ll}+\Delta \widetilde H ].c_l = \cE_{\rm RR} c_l \, ,  \label{heff2}
\ee
where $\Delta \widetilde H $ is given by
\be
 \Delta H_2\frac{1}{\Delta H_2 -\Delta H_3+(\cE_{\rm RR}-\cE_*)G} \Delta H_2 \, .  
\ee
In our calculations we will have $\cE_* \approx \cE_{\rm RR}$ \footnote{In practice we fix $\cE_*$ to the value given by the local approximation mentioned below Eq.~\reef{seriesdh}. Further iterative improvements are possible, but their effect is negligible.}. 
So we will neglect the last term in the denominator and will use
\be{
\Delta \widetilde H= \Delta H_2 \frac{1}{\Delta H_2-\Delta H_3} \Delta H_2}\, . \label{mainresult}
\ee
Now observe that the power expansion of this expression agrees, up to third order in $V$, with \reef{series1}:
\be
 \Delta \widetilde H = \Delta H_2 + \Delta H_3 + \dots \, . \label{exp3}
\ee
This key observation reveals the connection of the discussed method with the renormalization
idea from the previous section. Although this was not obvious from the start, Eq.~\reef{exp3} means that $\Delta \widetilde H$ implements a next-to-leading (NLO) renormalization correction. The presence of $\Delta H_3$ in the denominator of 
\eqref{exp3} is crucial to address the problems originating from the naive truncation of the expansion \eqref{seriesdh}. 
\footnote{By applying the power counting arguments mentioned in footonote 21, one can estimate $\Delta \tilde{H} \sim N E_T$, 
as opposed to $\Delta H_2 \sim N^2 E_T$, therefore taming the growth of the matrix elements.}
We will refer to the spectrum obtained via this method as NLO-HT.

{\bf Testing NLO-HT in the $(\phi^4)_2$ theory.} 
In the rest of the paper we will apply NLO-HT to one particular strongly coupled relativistic 
QFT---the $\phi^4$ theory in 1+1 dimensions. We stress however that the basic ideas of NLO-HT and of its implementation described below are general and can be used for many other theories.

We introduce here the $(\phi^4)_2$ theory very briefly; see \cite{Lorenzo1,JSL2} for details. The theory is defined by the normal-ordered Euclidean action 
\beq
S=\half \int d\tau\,dx\,[ \NO{(\del\phi)^2+m^2\phi^2}+g\, \NO{\phi^4}]\,.
\eeq
We quantize it canonically with periodic boundary conditions, expanding the field into creation and annihilation operators:
\beq
\phi(x,\tau=0)=\sum_k \frac{1}{\sqrt{2L\omega_k}}(a_k e^{ikx}+a_k^\dagger e^{-ikx}) \, ,
\eeq
where 
$ k=2\pi n/L\ (n\in \bZ)$, $\omega_k=\sqrt{m^2+k^2}$, $[a_k,a_{k'}]=0$ and $[a_k,a^\dagger_{k'}]=\delta_{kk'}$. Here $x$ is the coordinate along the spacial circle of length $L$, while $\tau\in\bR$ is the Euclidean time.
From now on, we will use the units $m=1$.

In terms of normal-ordered operators, the Hamiltonian is 
a sum of the free piece and the quartic interaction:
\begin{gather}
H=H_0 + g V_4 + \ldots
\label{eq:hamphi4}
\,,\qquad H_0=\sum_k \omega_k a^\dagger_k  a_k\,,\\
V_4 = L \sum_{\sum k_i=0} \frac 1{\prod\sqrt{2L\omega_i}}
\Big[ a_{k_1}a_{k_2}a_{k_3}a_{k_4} + \ldots \Big]\,.
\end{gather}
The ellipsis in $H$ in \reef{eq:hamphi4} refer to the Casimir energy and other exponentially suppressed corrections 
needed to correctly put the theory in finite volume. They are discussed
in detail in \cite{Lorenzo1} and defined in Eqs.~(2.10, 2.18) of that paper.
The Hamiltonian $H$ acts in the free theory Fock space. 
There are three conserved quantum numbers: total momentum $P$, spatial parity $\bP$ 
($x\to-x)$, and field parity $\bZ_2$ ($\phi\to -\phi$). 
We will focus on the invariant subspaces $\calH^{\pm}$ consisting of states with $P=0$,
$\bP=+$, $\bZ_2=\pm $. The states in $\calH^{+}$ (resp.~$\calH^{-}$) contain even 
(resp.~odd) number of free quanta.
The basic problem is to find eigenstates of $H$ belonging to $\calH^{\pm}$.  The two
subspaces do not mix, and the diagonalization can be done separately.

Let's describe briefly how the matrices entering the NLO-HT eigenvalue equation \reef{heff2} are computed in practice. The matrix elements of $H_{ll}$ are known in closed form and are straightforward to evaluate, taking advantage of the sparsity for efficiency. The matrices $\Delta H_{2,3}$ in \reef{mainresult} involve infinite sums over states in ${\cal H}_h$.  We approximate $\Delta H_2$ to high accuracy by 
splitting it as \cite{Elias-Miro:2015bqk}
\be
\Delta H_2= \Delta H_2^{<} + \Delta H_2^{>} \, .
\label{dh2approx}
\ee
Here 
the matrix $\Delta H_2^{<}$ involves a finite sum over the states in ${\cal H}_h$ of energies $E_T < E_i\leq E_L$ which is evaluated exactly. On the other hand, the matrix $\Delta H_2^{>}$ involves an infinite sum over
 the states with
 $E_i> E_L$, for which we use a local approximation
\cite{Hogervorst:2014rta,Lorenzo1}:
\be
\Delta H_2^> \approx \sum_{i} \kappa_i(E_L) \int_0^L dx\, \cO_i(x) \, .  \label{genloc}
\ee
Here the $\cO_i$ are a finite number
of local Lorentz-invariant operators; for the $(\phi^4)_2$ theory these are $\mathds{1}$, $\NO{\phi^2}$, $\NO{\phi^4}$. The coefficients $\kappa_i(E_L)$ are known analytically. 
This approximation is most accurate for matrix elements  
$(\Delta H_2)_{i j}$ such that $E_i, E_j \ll E_L$. Its validity is justified by the operator product 
expansion. 

 The original local approximation in Eq.~\reef{H2local} was given by the same formula \reef{genloc} but with $E_L=E_T$. So it was not accurate for states close to the cutoff. Instead, the error in evaluating $\Delta H_2$ via \reef{dh2approx} can be made arbitrarily small throughout the low-energy Hilbert space $\calH_l$ by raising $E_L$ above $E_T$.
In our calculations we find that $E_L=3E_T$ provides a sufficient approximation. The error can also be further reduced by including subleading (higher derivative) operators in \reef{genloc}. 

\begin{figure}[t]
\begin{center}
\includegraphics[scale=0.42]{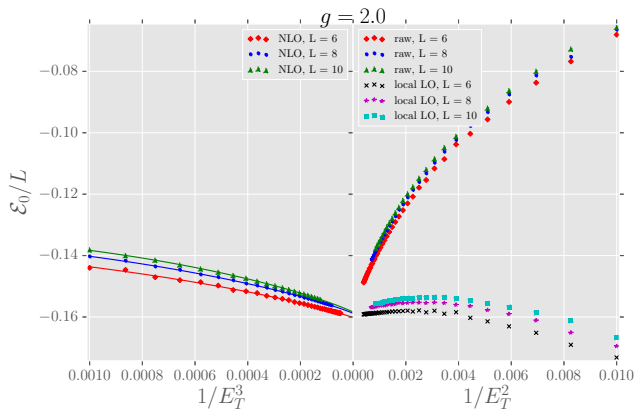} 
\caption{Convergence of NLO-HT vs other HT methods.}  \label{fig:ET}
\end{center}
\vspace{-0.5cm}
\end{figure}

The strategy for computing $\Delta H_3$ is analogous. We break down the matrix 
into various contributions. Some of those involve a finite sum over elements in 
$\calH_h$ close to the cutoff and are computed exactly. The remaining pieces contain the contributions 
of the states much above the cutoff. Those are approximated by a 
sum of local operators, with analytically known coefficients \cite{JSL2}.

{\bf Numerical results.}   The basic features of the low-lying $\phi^4$ spectrum are as follows. The lowest eigenstate $\calE_0$ belongs to $\calH^+$ and is the ground state in finite volume (the interacting vacuum).
 The second-lowest eigenstate belongs to $\calH^{-}$ and is interpreted as the one-particle 
excitation at zero momentum. The excitation energy over the ground state 
$\calE_1 - \calE_0$ measures the physical particle mass $m_{\rm ph}$. The above is true for moderate quartic coupling $g  < g_c \approx 2.8$, when the vacuum preserves the $\bZ_2$ invariance.  
At $g = g_c$ the particle mass goes to zero and the theory undergoes a second order phase transition to the phase of
spontaneously broken $\bZ_2$ symmetry, with critical
exponents given by the 2d critical Ising model.

We will now use the NLO-HT method to provide accurate non-perturbative predictions
for $\calE_0$ and $m_{\phys}$ as functions of the coupling $g$. Notice that perturbation theory ceases to be accurate for  $g\gtrsim 0.2$ (\cite{Lorenzo1}, appendix B). We will only study here the $\bZ_2$-invariant phase. The $\bZ_2$-broken phase at $g>g_c$ was studied previously in \cite{Coser:2014lla,Lorenzo2, Bajnok:2015bgw}.

While here we will focus on the vacuum and the first excited state, we stress that higher excited states and other observables are both possible and interesting to study using the HT. E.g.~one can extract the S-matrix from the volume dependence of the {two-particle} state energies \cite{Yurov:1991my}.

The first step is to compute the spectrum as a function of $E_T$ for fixed $g$ and $L$ and to extrapolate $E_T\to\infty$. 
Our NLO-HT calculations explored the couplings $g\le 3$ and the volumes $L\le 10$, while $E_T$ was fixed for each $L$ to have about $10^4$ states in $\calH_l$. 
For comparison, we will also report raw and local LO renormalized HT calculations, which were pushed to much higher $E_T$, corresponding to about $10^6$ states. As an indication of the needed computer resources, our most expensive NLO-HT data points ($L=10$, $E_T = 20$) required 40 CPU hours and 80 Gb RAM per coupling value.

Empirically, the NLO-HT spectrum was observed to converge with cutoff as $1/E_T^3$. A representative plot, for the vacuum energy at $g=2$, is in Fig.~\ref{fig:ET}(left). This is much faster than the raw and the local LO renormalized HT predictions for the same observable, which show $\sim 1/E_T^2$ convergence, although LO renormalization reduces the prefactor significantly, Fig.~\ref{fig:ET}(right). {The} smooth behavior of the NLO-HT data with $E_T$ allows us to extrapolate to $E_T=\infty$. For this we fit the NLO-HT data points with 
the function $F(E_T)=\alpha + \beta/E_T^3 + \gamma/E_T^4$,
with $\alpha$, $\beta$ and $\gamma$ free parameters, 
and use $F(\infty) =\alpha$ \footnote{To estimate extrapolation errors, we fitted subsamples of the full data set, obtained by removing points at low $E_T$ in different combinations.}. 

\begin{figure}[t]
\begin{center}
\vspace{-0.2cm}
\includegraphics[scale=.47]{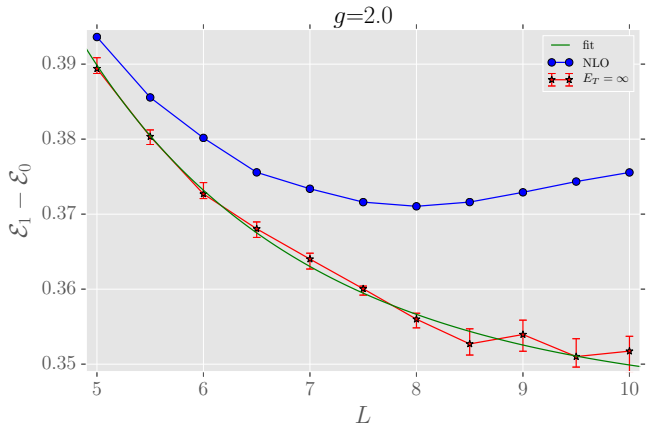}

\vspace{-.2cm}
\caption{The mass gap $\calE_1(L)-\calE_0(L)$ as a function of $L$.}  \label{fig:L}
\end{center}
\vspace{-0.54cm}
\end{figure}

Next we discuss how the spectrum depends on $L$. There are precise theoretical expectations for this dependence, which allows  us to perform interesting consistency checks,  
and helps to extrapolate  the mass gap and the vacuum energy
density to their infinite volume limits (for $g$ not too close to $g_c$).
For the
mass gap at $L m_\phys \gg 1$ we expect, in a 1+1 dimensional QFT with unbroken $\bZ_2$ symmetry
 \cite{Luscher:1985dn,Klassen:1990ub}:
\begin{gather}
\cE_1(L) - \cE_0(L) = m_\phys + \Delta m(L) + O(e^{- \sigma\, m_\phys L}) \,,  \label{eq:massfinL}
\\
\hspace{-0.2cm}\Delta m(L) = -\frac{1}{8 \pi m_\phys} \int d\theta \, e^{- m_\phys L 
\cosh \theta } F(\theta + i \pi/2)\,, \label{dM}
\\  
F(\theta) = -4 i m_\phys^2 \sinh(\theta) \left(S(\theta) -1 \right)\,,
\end{gather}
where $\sigma \ge \sqrt{3}$, and $S(\theta)$ is the S-matrix for $2 \to 2$ scattering,
with $\theta$ the rapidity difference. 

We neglect the third term in the r.h.s.~of \reef{eq:massfinL}, while we approximate the second one as follows. In this work we will not measure the S-matrix, 
\footnote{As a further check of the method, the S-matrix (extracted via the volume
dependence of the spectrum) could in the future be compared to the 
perturbative prediction for $g \ll 1$. See \cite{Coser:2014lla}, appendix B.}
but we will instead parametrize it by replacing $S(\theta + i \pi/2)$ with a series expansion 
around $\theta=0$. This is reasonable because the integral in $\Delta m(L)$ 
is dominated by small $\theta$. 
Eq.~\reef{dM} then implies:
\begin{eqnarray}
\Delta m(L)/m_{\phys} \approx b K_1(m_\phys L) + \frac{c}{(m_\phys L)^{3/2}} e^{-{L m_{\phys}}}\,. \quad \quad 
\label{eq:Lfit}
\end{eqnarray}
The Bessel function comes from the constant term of the 
$S(\theta)$ expansion, while the second term comes from  
 doing the integral via the steepest descent of the $\theta^2$ term
(the linear term vanishes in the integral). 
Further corrections are suppressed by additional powers of $m_{\phys} L$.

\begin{figure}[h]
\begin{center}
\includegraphics[scale=.42]{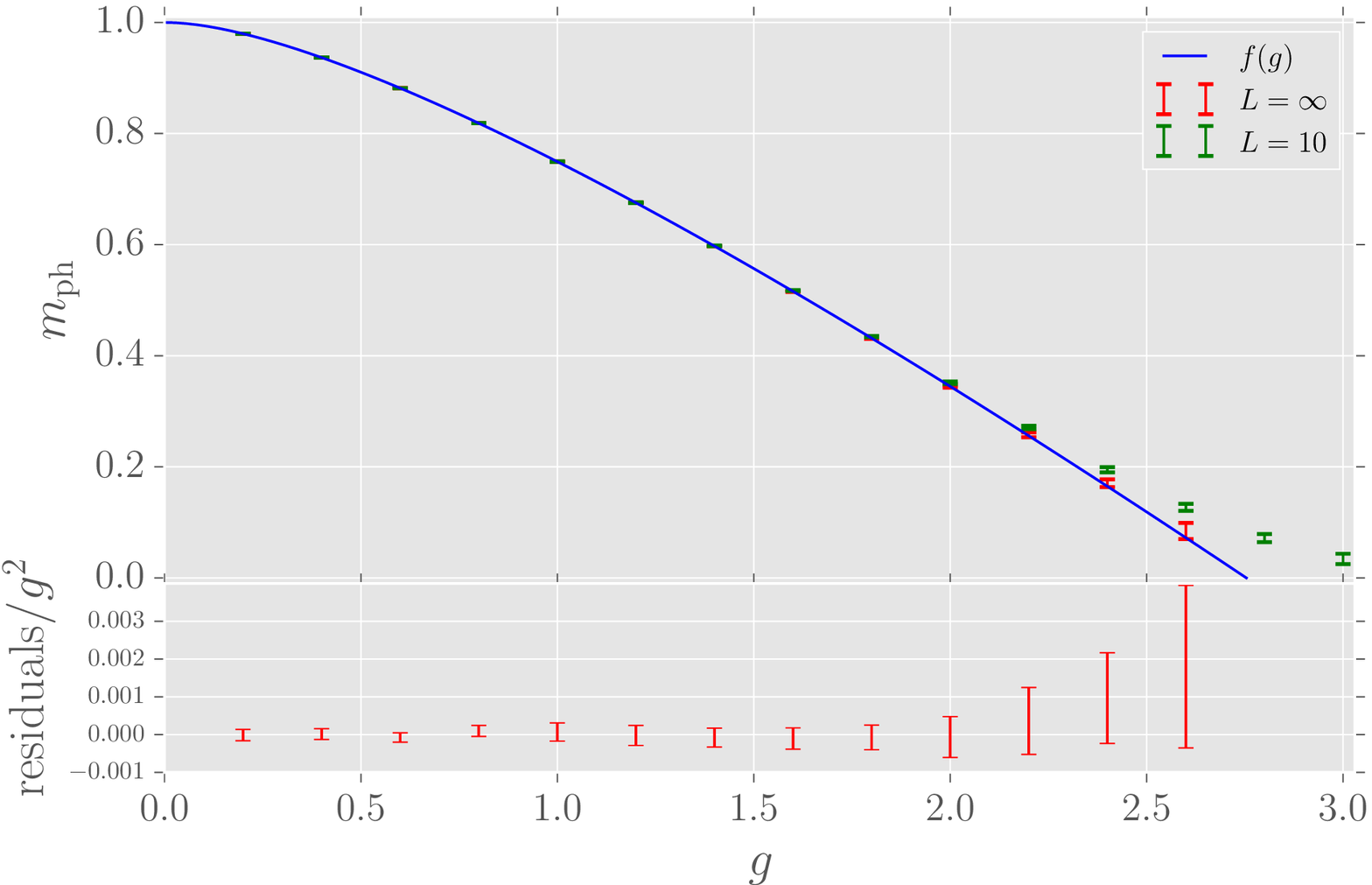} 
\vspace{-0.2cm}
\caption{$m_{\phys}(g)$; compare with Fig.~11 in \cite{Lorenzo1}.}
\label{fig:specvsG}
\end{center}
\vspace{-0.54cm}
\end{figure}

In Fig.~\ref{fig:L} the above expectations are compared to the $g=2$ NLO-HT data. We include the NLO-HT data points at the highest $E_T$ we could reach for the given $L$ (blue), and the NLO-HT data extrapolated
to $E_T=\infty$ as discussed above (red error bars). We also include the fit of the extrapolated data using Eq.~\reef{eq:Lfit} (green curve). The fit has three parameters ($m_\phys$, $b$, $c$) and works well in the
whole range of $L$. We extract the value of $m_\phys$ at $L\rightarrow \infty$ from the fit, with
the uncertainty determined by fitting the upper and lower ends of the error bars.
We have done analogous $L\rightarrow \infty$ extrapolations for all couplings $g\le 2.6$
in steps of $0.2$. These are shown in Fig.~\ref{fig:specvsG} (red error bars), where the $L=10$ results extrapolated to $E_T=\infty$ are also shown for comparison (green error bars). A few $L=\infty$ results are also reported in Table \ref{table:numerics}. For $g>2.6$, close to the critical point, the described fitting procedure cannot be used, as the physical mass approaches zero, and the condition $L m_\phys \gg 1$ is not satisfied. 

\begin{table}[t]
\centering
\begin{tabular}{L{1cm}  L{2.5cm}  L{2.5cm} }
$g$ & $m_\text{ph}$ & $\Lambda$ \\
\hline
0.2 & 0.979733(5) & $-0.0018166(5)$\\
1  & 0.7494(2) &  $-0.03941(2)$\\
2 & 0.345(2) & $-0.1581(1)$\\
\end{tabular}
\caption{$m_\text{ph}$ and $\Lambda$ extracted with NLO-HT.}
\label{table:numerics}
\vspace{-.35cm}
\end{table}

Also in Table \ref{table:numerics}, we report analogous measurements of the infinite volume vacuum energy density $\Lambda$ (the cosmological constant). The NLO-HT data for $\calE_0(L)/L$ are extrapolated to $E_T=\infty$ and then are fitted with the theoretical expectation
at $L m_{\phys}\gg 1$:
\be
\frac{\calE_0(L)}{L} = \Lambda -\frac{m_\phys}{\pi L} K_1(m_\phys L)
+ a \sqrt{\frac{m_\phys}{16\pi L^3}}e^{-2 m_\phys L} +\ldots,   \nonumber
\ee
where $a=O(1)$.
This formula is valid in any massive quantum field theory in 1+1 dimensions in
absence of bound states \cite{Klassen:1990dx, Lorenzo1}. 

Coming back to Fig.~\ref{fig:specvsG}, we see by eye that the mass gap vanishes somewhere close to $g_c\approx 2.8$, signaling a quantum critical point. This is in accord with previous theoretical \cite{Chang:1976ek} and numerical \cite{Schaich:2009jk,Lorenzo1,Bajnok:2015bgw,Milsted:2013rxa,Bosetti:2015lsa,Pelissetto:2015yha} 
studies \footnote{To compare with the critical coupling extractions using the light front quantization \cite{Chabysheva:2015ynr,Burkardt:2016ffk,Anand:2017yij} one has to perform nonperturbative mass renormalization \cite{Burkardt:2016ffk}.}. 
For a better estimate of $g_c$, we fit the $L=\infty$ data points
in the range $g\le 2.6$ with the rational function 
\be
f(g) = \frac{( 1 + g (\frac{1}{g_1} + \frac{1}{g_2} + \frac{1}{g_3} + \frac{1}{g_c}) 
+ r g^2) (1 - \frac{g}{g_c})^\nu}
{( 1 + \frac{g}{g_1}) (1 + \frac{g}{g_2}) (1 + \frac{g}{g_3}) } \, , 
\ee
with fit parameters $r$, $g_1$, $g_2$, $g_3$, $g_c$, and $\nu$. We have $f(g_c)=0$ by construction. We impose $g_1,g_2,g_3>0$ so that $f(g)$ has poles on the negative real axis. The critical coupling estimate from this fit 
is \footnote{The central value corresponds to the smallest $\chi^2(g_c)= \sum_{i=1}^N {(y_i-f(x_i))^2}/{err_i}^2$. The uncertainty interval was conservatively determined from the condition $\sqrt{\chi^2(g_c)}\le 3\sqrt{\chi^2(2.76)}$. Our determination is the best HT measurement of $g_c$. It is compatible with and has accuracy comparable to other available determinations \cite{JSL2}.}
 \be
g_c = 2.76(3)\,. \label{eq:gc}
\ee
The $\nu$ parameter in the above fit is a critical exponent. Assuming the Ising model universality class for the phase transition, we expect $\nu = (2 - \Delta_\epsilon)^{-1} = 1$, using  $\Delta_\epsilon = 1$, the dimension of the most relevant non-trivial $\bZ_2$-even operator of the critical Ising model. In the fit leading to \reef{eq:gc} we fixed $\nu=1$. Relaxing this assumption gives the same central value with slightly larger error bars.

The rationale behind introducing the poles into the ansatz $f(g)$ is that they are
supposed to approximate the branch cut at $g<0$ that the analytically continued function 
$m_{\phys} (g)$ is expected to have. 
We checked that modifying our ansatz, and in particular increasing the number
of poles, does not affect appreciably the confidence interval for $g_c$. 
We also checked that the $g^2$ and $g^3$ coefficients of our best fit are roughly
consistent with the perturbation theory prediction 
$
m_{\phys}(g)=1-1.5 g^2+2.86460(20) g^3+\ldots
$ \cite{Lorenzo1}.
With a more complicated ansatz, we found fits perfectly agreeing with perturbation theory. The resulting $g_c$ values are nearly identical to \reef{eq:gc}. This is not surprising, since most of fit power relevant for
constraining $g_c$ comes from $1\lesssim g\lesssim 2$, not from the region of small $g$ where perturbation
theory is accurate.
 
\begin{figure}[t]
\begin{center}
 \includegraphics[scale=.43]{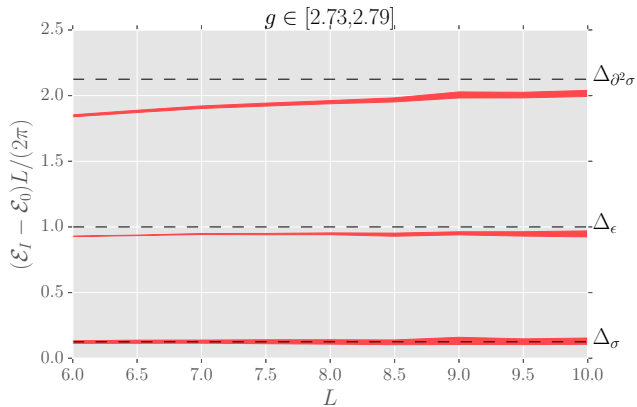} 
 \vspace{-0.2cm}
\caption{Energy levels at $g=g_c$ vs CFT predictions.}
\label{fig:specvsG1}
\end{center}
\vspace{-0.5cm}
\end{figure}

Finally, we compare the NLO-HT results to the expectations for the finite volume spectra at the critical point. CFT predicts that the energy levels at $g=g_c$ 
should vary with $L$ as
\vspace{-0.2cm}
\begin{equation}
\cE_I(L) - \cE_0(L) \approx 2\pi \Delta_I/L\,,
\label{eq:cft}
\end{equation}
where $\Delta_I$ are operator dimensions in the critical Ising model. This relation should hold at $L\gg 1$, where corrections due to irrelevant couplings die out.
In Fig.~\ref{fig:specvsG1} we test it for the first three 
energy levels above the vacuum, which should correspond to the operators 
with dimensions $\Delta_\sigma = 1/8$, $\Delta_\epsilon = 1$, 
$\Delta_{\partial^2 \sigma} = 2 + 1/8$. The error comes from extrapolating to $E_T=\infty$ and (the largest contribution) from varying $g$ in the range \reef{eq:gc}.
We see reasonable agreement for $\sigma$ and $\epsilon$, 
while it looks like the agreement for $\partial^2 \sigma$ 
will be reached at higher values of $L$. This figure can be compared to Fig.~6
of \cite{Lorenzo1} and Figs.~22, 23 of \cite{Bajnok:2015bgw}, which show 
similar behavior.

{\bf Conclusions.}
In this work we proposed a variant of renormalized Hamiltonian Truncation called NLO-HT. Its main idea is to integrate out exactly a certain class of high-energy states, which allows for variational interpretation, and furthermore implements the renormalization corrections up to cubic order in the interaction strength. 

We tested NLO-HT by computing the low-lying spectra of the strongly coupled two-dimensional $\phi^4$ theory. 
Numerical spectra in finite volume were found to converge rapidly with the Hilbert space cutoff $E_T$, faster than for other existing versions of Hamiltonian Truncation, and allowing controlled extrapolation to the continuum limit $E_T=\infty$. The finite volume corrections were then removed using the theoretical knowledge of 
these effects in QFT. In this way we extracted highly accurate 
predictions for the vacuum energy density and the physical mass in the infinite volume limit, for a range of non-perturbative coupling constants.

In the future NLO-HT will be used to perform accurate studies in other strongly coupled RG flows in $d=2$.
In particular, it can be applied to flows starting from an interacting CFTs. We also believe that our ideas will be useful to extend Hamiltonian Truncation to weakly relevant interactions, with scaling dimension in the range $\Delta_V>d/2$ excluded in this paper, and in particular to flows in higher dimensions $d\geq 3$, most of which fall into this category.

\vspace{5pt}

\begin{acknowledgments}
We thank Richard Brower, Ami Katz, Robert Konik, Iman Mahyaeh, Marco Serone, Gabor Tak\'acs, Giovanni Villadoro and Matthew Walters for the useful discussions. SR is supported by the National Centre of Competence in Research SwissMAP funded by the Swiss National Science Foundation, and by the Simons Foundation grant 488655 (Simons collaboration on the Non-perturbative bootstrap). 
The work of LV was supported by the Simons Foundation grant on the Nonperturbative 
Bootstrap and by the Swiss National Science Foundation under grant 200020-150060.
The computations were performed on the BU SCC  and SISSA Ulysses clusters. 
\end{acknowledgments}

\vspace{-.3cm}

\bibliography{phi4-Biblio2}

\begin{thebibliography}{47}%
\makeatletter
\providecommand \@ifxundefined [1]{%
 \@ifx{#1\undefined}
}%
\providecommand \@ifnum [1]{%
 \ifnum #1\expandafter \@firstoftwo
 \else \expandafter \@secondoftwo
 \fi
}%
\providecommand \@ifx [1]{%
 \ifx #1\expandafter \@firstoftwo
 \else \expandafter \@secondoftwo
 \fi
}%
\providecommand \natexlab [1]{#1}%
\providecommand \enquote  [1]{``#1''}%
\providecommand \bibnamefont  [1]{#1}%
\providecommand \bibfnamefont [1]{#1}%
\providecommand \citenamefont [1]{#1}%
\providecommand \href@noop [0]{\@secondoftwo}%
\providecommand \href [0]{\begingroup \@sanitize@url \@href}%
\providecommand \@href[1]{\@@startlink{#1}\@@href}%
\providecommand \@@href[1]{\endgroup#1\@@endlink}%
\providecommand \@sanitize@url [0]{\catcode `\\12\catcode `\$12\catcode
  `\&12\catcode `\#12\catcode `\^12\catcode `\_12\catcode `\%12\relax}%
\providecommand \@@startlink[1]{}%
\providecommand \@@endlink[0]{}%
\providecommand \url  [0]{\begingroup\@sanitize@url \@url }%
\providecommand \@url [1]{\endgroup\@href {#1}{\urlprefix }}%
\providecommand \urlprefix  [0]{URL }%
\providecommand \Eprint [0]{\href }%
\providecommand \doibase [0]{http://dx.doi.org/}%
\providecommand \selectlanguage [0]{\@gobble}%
\providecommand \bibinfo  [0]{\@secondoftwo}%
\providecommand \bibfield  [0]{\@secondoftwo}%
\providecommand \translation [1]{[#1]}%
\providecommand \BibitemOpen [0]{}%
\providecommand \bibitemStop [0]{}%
\providecommand \bibitemNoStop [0]{.\EOS\space}%
\providecommand \EOS [0]{\spacefactor3000\relax}%
\providecommand \BibitemShut  [1]{\csname bibitem#1\endcsname}%
\let\auto@bib@innerbib\@empty
\bibitem [{\citenamefont {White}(1993)}]{White:1993zza}%
  \BibitemOpen
  \bibfield  {author} {\bibinfo {author} {\bibfnamefont {S.~R.}\ \bibnamefont
  {White}},\ }\bibfield  {title} {\enquote {\bibinfo {title} {{Density-matrix
  algorithms for quantum renormalization groups}},}\ }\href {\doibase
  10.1103/PhysRevB.48.10345} {\bibfield  {journal} {\bibinfo  {journal} {Phys.
  Rev.}\ }\textbf {\bibinfo {volume} {B48}},\ \bibinfo {pages} {10345}
  (\bibinfo {year} {1993})}\BibitemShut {NoStop}%
\bibitem [{\citenamefont {{Perez-Garcia}}\ \emph {et~al.}(2007)\citenamefont
  {{Perez-Garcia}}, \citenamefont {{Verstraete}}, \citenamefont {{Wolf}},\ and\
  \citenamefont {{Cirac}}}]{MPSreps}%
  \BibitemOpen
  \bibfield  {author} {\bibinfo {author} {\bibfnamefont {D.}~\bibnamefont
  {{Perez-Garcia}}}, \bibinfo {author} {\bibfnamefont {F.}~\bibnamefont
  {{Verstraete}}}, \bibinfo {author} {\bibfnamefont {M.~M.}\ \bibnamefont
  {{Wolf}}}, \ and\ \bibinfo {author} {\bibfnamefont {J.~I.}\ \bibnamefont
  {{Cirac}}},\ }\bibfield  {title} {\enquote {\bibinfo {title} {{Matrix Product
  State Representations}},}\ }\href@noop {} {\bibfield  {journal} {\bibinfo
  {journal} {Quantum Inf. Comput.}\ }\textbf {\bibinfo {volume} {7}},\ \bibinfo
  {pages} {401} (\bibinfo {year} {2007})},\ \Eprint
  {http://arxiv.org/abs/quant-ph/0608197} {quant-ph/0608197} \BibitemShut
  {NoStop}%
\bibitem [{\citenamefont {{Shi}}\ \emph {et~al.}(2006)\citenamefont {{Shi}},
  \citenamefont {{Duan}},\ and\ \citenamefont {{Vidal}}}]{ShiDuanVidal}%
  \BibitemOpen
  \bibfield  {author} {\bibinfo {author} {\bibfnamefont {Y.-Y.}\ \bibnamefont
  {{Shi}}}, \bibinfo {author} {\bibfnamefont {L.-M.}\ \bibnamefont {{Duan}}}, \
  and\ \bibinfo {author} {\bibfnamefont {G.}~\bibnamefont {{Vidal}}},\
  }\bibfield  {title} {\enquote {\bibinfo {title} {{Classical simulation of
  quantum many-body systems with a tree tensor network}},}\ }\href {\doibase
  10.1103/PhysRevA.74.022320} {\bibfield  {journal} {\bibinfo  {journal}
  {Phys.~Rev.}\ }\textbf {\bibinfo {volume} {A74}},\ \bibinfo {pages} {022320}
  (\bibinfo {year} {2006})},\ \Eprint {http://arxiv.org/abs/quant-ph/0511070}
  {quant-ph/0511070} \BibitemShut {NoStop}%
\bibitem [{\citenamefont {{Vidal}}(2008)}]{MERA}%
  \BibitemOpen
  \bibfield  {author} {\bibinfo {author} {\bibfnamefont {G.}~\bibnamefont
  {{Vidal}}},\ }\bibfield  {title} {\enquote {\bibinfo {title} {{Class of
  Quantum Many-Body States That Can Be Efficiently Simulated}},}\ }\href
  {\doibase 10.1103/PhysRevLett.101.110501} {\bibfield  {journal} {\bibinfo
  {journal} {Phys.~Rev.~Lett.}\ }\textbf {\bibinfo {volume} {101}},\ \bibinfo
  {pages} {110501} (\bibinfo {year} {2008})},\ \Eprint
  {http://arxiv.org/abs/quant-ph/0610099} {quant-ph/0610099} \BibitemShut
  {NoStop}%
\bibitem [{\citenamefont {{Verstraete}}\ \emph {et~al.}(2008)\citenamefont
  {{Verstraete}}, \citenamefont {{Murg}},\ and\ \citenamefont
  {{Cirac}}}]{PEPS}%
  \BibitemOpen
  \bibfield  {author} {\bibinfo {author} {\bibfnamefont {F.}~\bibnamefont
  {{Verstraete}}}, \bibinfo {author} {\bibfnamefont {V.}~\bibnamefont
  {{Murg}}}, \ and\ \bibinfo {author} {\bibfnamefont {J.~I.}\ \bibnamefont
  {{Cirac}}},\ }\bibfield  {title} {\enquote {\bibinfo {title} {{Matrix product
  states, projected entangled pair states, and variational renormalization
  group methods for quantum spin systems}},}\ }\href {\doibase
  10.1080/14789940801912366} {\bibfield  {journal} {\bibinfo  {journal}
  {Advances in Physics}\ }\textbf {\bibinfo {volume} {57}},\ \bibinfo {pages}
  {143--224} (\bibinfo {year} {2008})},\ \Eprint
  {http://arxiv.org/abs/0907.2796} {arXiv:0907.2796 [quant-ph]} \BibitemShut
  {NoStop}%
\bibitem [{\citenamefont {Yurov}\ and\ \citenamefont
  {Zamolodchikov}(1990)}]{Yurov:1989yu}%
  \BibitemOpen
  \bibfield  {author} {\bibinfo {author} {\bibfnamefont {V.~P.}\ \bibnamefont
  {Yurov}}\ and\ \bibinfo {author} {\bibfnamefont {Al.~B.}\ \bibnamefont
  {Zamolodchikov}},\ }\bibfield  {title} {\enquote {\bibinfo {title}
  {{Truncated Conformal Space Approach to Scaling Lee-Yang Model}},}\ }\href
  {\doibase 10.1142/S0217751X9000218X} {\bibfield  {journal} {\bibinfo
  {journal} {Int.J.Mod.Phys.}\ }\textbf {\bibinfo {volume} {A5}},\ \bibinfo
  {pages} {3221--3246} (\bibinfo {year} {1990})}\BibitemShut {NoStop}%
\bibitem [{\citenamefont {Yurov}\ and\ \citenamefont
  {Zamolodchikov}(1991)}]{Yurov:1991my}%
  \BibitemOpen
  \bibfield  {author} {\bibinfo {author} {\bibfnamefont {V.~P.}\ \bibnamefont
  {Yurov}}\ and\ \bibinfo {author} {\bibfnamefont {Al.~B.}\ \bibnamefont
  {Zamolodchikov}},\ }\bibfield  {title} {\enquote {\bibinfo {title}
  {{Truncated fermionic space approach to the critical 2-D Ising model with
  magnetic field}},}\ }\href {\doibase 10.1142/S0217751X91002161} {\bibfield
  {journal} {\bibinfo  {journal} {Int.J.Mod.Phys.}\ }\textbf {\bibinfo {volume}
  {A6}},\ \bibinfo {pages} {4557} (\bibinfo {year} {1991})}\BibitemShut
  {NoStop}%
\bibitem [{\citenamefont {{James}}\ \emph {et~al.}()\citenamefont {{James}},
  \citenamefont {{Konik}}, \citenamefont {{Lecheminant}}, \citenamefont
  {{Robinson}},\ and\ \citenamefont {{Tsvelik}}}]{Konik-review}%
  \BibitemOpen
  \bibfield  {author} {\bibinfo {author} {\bibfnamefont {A.~J.~A.}\
  \bibnamefont {{James}}}, \bibinfo {author} {\bibfnamefont {R.~M.}\
  \bibnamefont {{Konik}}}, \bibinfo {author} {\bibfnamefont {P.}~\bibnamefont
  {{Lecheminant}}}, \bibinfo {author} {\bibfnamefont {N.~J.}\ \bibnamefont
  {{Robinson}}}, \ and\ \bibinfo {author} {\bibfnamefont {A.~M.}\ \bibnamefont
  {{Tsvelik}}},\ }\bibfield  {title} {\enquote {\bibinfo {title}
  {{Non-perturbative methodologies for low-dimensional strongly-correlated
  systems: From non-abelian bosonization to truncated spectrum methods}},}\
  }\href@noop {} {\ }\Eprint {http://arxiv.org/abs/1703.08421}
  {arXiv:1703.08421 [cond-mat.str-el]} \BibitemShut {NoStop}%
\bibitem [{Note1()}]{Note1}%
  \BibitemOpen
  \bibinfo {note} {In relativistic QFTs one can also quantize on surfaces of
  constant light-cone coordinate. This \protect \textit {light front
  quantization} \cite {Brodsky:1997de} is also used in numerical solutions of
  strongly coupled QFTs via a version of HT; some recent work is \cite
  {Katz:2013qua,Katz:2014uoa,Chabysheva:2015ynr,Burkardt:2016ffk,Katz:2016hxp,Anand:2017yij}.
  The structure of the unperturbed Hilbert space is different from the equal
  time case, which leads to important differences in the numerical procedure.
  All technical claims in this work will refer exclusively to the equal time
  quantization.}\BibitemShut {Stop}%
\bibitem [{\citenamefont {Klassen}\ and\ \citenamefont
  {Melzer}(1992)}]{Klassen:1991ze}%
  \BibitemOpen
  \bibfield  {author} {\bibinfo {author} {\bibfnamefont {T.~R.}\ \bibnamefont
  {Klassen}}\ and\ \bibinfo {author} {\bibfnamefont {E.}~\bibnamefont
  {Melzer}},\ }\bibfield  {title} {\enquote {\bibinfo {title} {{Spectral flow
  between conformal field theories in (1+1) dimensions}},}\ }\href {\doibase
  10.1016/0550-3213(92)90422-8} {\bibfield  {journal} {\bibinfo  {journal}
  {Nucl.Phys.}\ }\textbf {\bibinfo {volume} {B370}},\ \bibinfo {pages}
  {511--550} (\bibinfo {year} {1992})}\BibitemShut {NoStop}%
\bibitem [{\citenamefont {Hogervorst}\ \emph {et~al.}(2015)\citenamefont
  {Hogervorst}, \citenamefont {Rychkov},\ and\ \citenamefont {van
  Rees}}]{Hogervorst:2014rta}%
  \BibitemOpen
  \bibfield  {author} {\bibinfo {author} {\bibfnamefont {M.}~\bibnamefont
  {Hogervorst}}, \bibinfo {author} {\bibfnamefont {S.}~\bibnamefont {Rychkov}},
  \ and\ \bibinfo {author} {\bibfnamefont {B.~C.}\ \bibnamefont {van Rees}},\
  }\bibfield  {title} {\enquote {\bibinfo {title} {{Truncated conformal space
  approach in $d$ dimensions: A cheap alternative to lattice field theory?}}}\
  }\href {\doibase 10.1103/PhysRevD.91.025005} {\bibfield  {journal} {\bibinfo
  {journal} {Phys. Rev.}\ }\textbf {\bibinfo {volume} {D91}},\ \bibinfo {pages}
  {025005} (\bibinfo {year} {2015})},\ \Eprint {http://arxiv.org/abs/1409.1581}
  {arXiv:1409.1581 [hep-th]} \BibitemShut {NoStop}%
\bibitem [{\citenamefont {Rychkov}\ and\ \citenamefont
  {Vitale}(2015)}]{Lorenzo1}%
  \BibitemOpen
  \bibfield  {author} {\bibinfo {author} {\bibfnamefont {S.}~\bibnamefont
  {Rychkov}}\ and\ \bibinfo {author} {\bibfnamefont {L.~G.}\ \bibnamefont
  {Vitale}},\ }\bibfield  {title} {\enquote {\bibinfo {title} {{Hamiltonian
  truncation study of the $\phi^4$ theory in two dimensions}},}\ }\href
  {\doibase 10.1103/PhysRevD.91.085011} {\bibfield  {journal} {\bibinfo
  {journal} {Phys. Rev.}\ }\textbf {\bibinfo {volume} {D91}},\ \bibinfo {pages}
  {085011} (\bibinfo {year} {2015})},\ \Eprint {http://arxiv.org/abs/1412.3460}
  {arXiv:1412.3460 [hep-th]} \BibitemShut {NoStop}%
\bibitem [{\citenamefont {Rychkov}\ and\ \citenamefont
  {Vitale}(2016)}]{Lorenzo2}%
  \BibitemOpen
  \bibfield  {author} {\bibinfo {author} {\bibfnamefont {S.}~\bibnamefont
  {Rychkov}}\ and\ \bibinfo {author} {\bibfnamefont {L.~G.}\ \bibnamefont
  {Vitale}},\ }\bibfield  {title} {\enquote {\bibinfo {title} {{Hamiltonian
  truncation study of the $\phi^4$ theory in two dimensions. II. The $\mathbb
  Z_2$ -broken phase and the Chang duality}},}\ }\href {\doibase
  10.1103/PhysRevD.93.065014} {\bibfield  {journal} {\bibinfo  {journal} {Phys.
  Rev.}\ }\textbf {\bibinfo {volume} {D93}},\ \bibinfo {pages} {065014}
  (\bibinfo {year} {2016})},\ \Eprint {http://arxiv.org/abs/1512.00493}
  {arXiv:1512.00493 [hep-th]} \BibitemShut {NoStop}%
\bibitem [{\citenamefont {Elias-Mir\'o}\ \emph {et~al.}(2016)\citenamefont
  {Elias-Mir\'o}, \citenamefont {Montull},\ and\ \citenamefont
  {Riembau}}]{Elias-Miro:2015bqk}%
  \BibitemOpen
  \bibfield  {author} {\bibinfo {author} {\bibfnamefont {J.}~\bibnamefont
  {Elias-Mir\'o}}, \bibinfo {author} {\bibfnamefont {M.}~\bibnamefont
  {Montull}}, \ and\ \bibinfo {author} {\bibfnamefont {M.}~\bibnamefont
  {Riembau}},\ }\bibfield  {title} {\enquote {\bibinfo {title} {{The
  renormalized Hamiltonian truncation method in the large $E_T$ expansion}},}\
  }\href {\doibase 10.1007/JHEP04(2016)144} {\bibfield  {journal} {\bibinfo
  {journal} {JHEP}\ }\textbf {\bibinfo {volume} {04}},\ \bibinfo {pages} {144}
  (\bibinfo {year} {2016})},\ \Eprint {http://arxiv.org/abs/1512.05746}
  {arXiv:1512.05746} \BibitemShut {NoStop}%
\bibitem [{\citenamefont {Giokas}\ and\ \citenamefont
  {Watts}(2011)}]{Giokas:2011ix}%
  \BibitemOpen
  \bibfield  {author} {\bibinfo {author} {\bibfnamefont {P.}~\bibnamefont
  {Giokas}}\ and\ \bibinfo {author} {\bibfnamefont {G.}~\bibnamefont {Watts}},\
  }\bibfield  {title} {\enquote {\bibinfo {title} {{The renormalisation group
  for the truncated conformal space approach on the cylinder}},}\ }\href@noop
  {} {\  (\bibinfo {year} {2011})},\ \Eprint {http://arxiv.org/abs/1106.2448}
  {arXiv:1106.2448 [hep-th]} \BibitemShut {NoStop}%
\bibitem [{\citenamefont {Feverati}\ \emph {et~al.}(2008)\citenamefont
  {Feverati}, \citenamefont {Graham}, \citenamefont {Pearce}, \citenamefont
  {Toth},\ and\ \citenamefont {Watts}}]{Feverati:2006ni}%
  \BibitemOpen
  \bibfield  {author} {\bibinfo {author} {\bibfnamefont {G.}~\bibnamefont
  {Feverati}}, \bibinfo {author} {\bibfnamefont {K.}~\bibnamefont {Graham}},
  \bibinfo {author} {\bibfnamefont {P.~A.}\ \bibnamefont {Pearce}}, \bibinfo
  {author} {\bibfnamefont {G.~Zs.}\ \bibnamefont {Toth}}, \ and\ \bibinfo
  {author} {\bibfnamefont {G.}~\bibnamefont {Watts}},\ }\bibfield  {title}
  {\enquote {\bibinfo {title} {{A Renormalisation group for TCSA}},}\ }\href
  {\doibase 10.1088/1742-5468/2008/03/P03011} {\bibfield  {journal} {\bibinfo
  {journal} {J. Stat. Mech.}\ ,\ \bibinfo {pages} {P03011}} (\bibinfo {year}
  {2008})},\ \Eprint {http://arxiv.org/abs/hep-th/0612203}
  {arXiv:hep-th/0612203 [hep-th]} \BibitemShut {NoStop}%
\bibitem [{\citenamefont {Watts}(2012)}]{Watts:2011cr}%
  \BibitemOpen
  \bibfield  {author} {\bibinfo {author} {\bibfnamefont {G.}~\bibnamefont
  {Watts}},\ }\bibfield  {title} {\enquote {\bibinfo {title} {{On the
  renormalisation group for the boundary Truncated Conformal Space
  Approach}},}\ }\href {\doibase 10.1016/j.nuclphysb.2012.01.012} {\bibfield
  {journal} {\bibinfo  {journal} {Nucl.Phys.}\ }\textbf {\bibinfo {volume}
  {B859}},\ \bibinfo {pages} {177--206} (\bibinfo {year} {2012})},\ \Eprint
  {http://arxiv.org/abs/1104.0225} {arXiv:1104.0225 [hep-th]} \BibitemShut
  {NoStop}%
\bibitem [{\citenamefont {Lencses}\ and\ \citenamefont
  {Takacs}(2014)}]{Lencses:2014tba}%
  \BibitemOpen
  \bibfield  {author} {\bibinfo {author} {\bibfnamefont {M.}~\bibnamefont
  {Lencses}}\ and\ \bibinfo {author} {\bibfnamefont {G.}~\bibnamefont
  {Takacs}},\ }\bibfield  {title} {\enquote {\bibinfo {title} {{Excited state
  TBA and renormalized TCSA in the scaling Potts model}},}\ }\href {\doibase
  10.1007/JHEP09(2014)052} {\bibfield  {journal} {\bibinfo  {journal} {JHEP}\
  }\textbf {\bibinfo {volume} {09}},\ \bibinfo {pages} {052} (\bibinfo {year}
  {2014})},\ \Eprint {http://arxiv.org/abs/1405.3157} {arXiv:1405.3157
  [hep-th]} \BibitemShut {NoStop}%
\bibitem [{\citenamefont {Elias-Miro}\ \emph {et~al.}(2017)\citenamefont
  {Elias-Miro}, \citenamefont {Rychkov},\ and\ \citenamefont {Vitale}}]{JSL2}%
  \BibitemOpen
  \bibfield  {author} {\bibinfo {author} {\bibfnamefont {Joan}\ \bibnamefont
  {Elias-Miro}}, \bibinfo {author} {\bibfnamefont {Slava}\ \bibnamefont
  {Rychkov}}, \ and\ \bibinfo {author} {\bibfnamefont {Lorenzo~G.}\
  \bibnamefont {Vitale}},\ }\bibfield  {title} {\enquote {\bibinfo {title}
  {{NLO Renormalization in the Hamiltonian Truncation}},}\ }\href {\doibase
  10.1103/PhysRevD.96.065024} {\bibfield  {journal} {\bibinfo  {journal} {Phys.
  Rev.}\ }\textbf {\bibinfo {volume} {D96}},\ \bibinfo {pages} {065024}
  (\bibinfo {year} {2017})},\ \Eprint {http://arxiv.org/abs/1706.09929}
  {arXiv:1706.09929 [hep-th]} \BibitemShut {NoStop}%
\bibitem [{Note2()}]{Note2}%
  \BibitemOpen
  \bibinfo {note} {This has to be distinguished from the Numerical
  Renormalization Group (NRG) improvement of the HT \cite {Konik:2007cb} \`a la
  Wilson's NRG \cite {Wilson}. This method raises the cutoff by adding new
  chunks of the Hilbert space and tossing away the states which have low
  overlaps with the interacting eigenstates. Other ideas to extend the reach of
  HT include \protect \textit {sweeping} and \protect \textit {reordering} (see
  \cite {Konik-review}). We have not used any of these interesting tricks in
  our work.}\BibitemShut {Stop}%
\bibitem [{Note3()}]{Note3}%
  \BibitemOpen
  \bibinfo {note} {That can be intuitively understood as follows. For each $n
  \geqslant 2$, there will be states below the cutoff for which the matrix
  elements of $\Delta H_n$ grow as $\sim (c N)^n E_T$ in absolute value, where
  $N$ is the occupation number of the state and $c$ is a constant. For $N$ big
  enough, the expansion is therefore not convergent, as the truncated matrix
  element will outgrow the leading order contribution of $H_0$ growing as $\sim
  E_T$. For a detailed discussion of this point see \cite {JSL2}, appendix
  B.}\BibitemShut {Stop}%
\bibitem [{Note4()}]{Note4}%
  \BibitemOpen
  \bibinfo {note} {In this work, we introduce a tail state for each $\ket
  {i}\in \calH _l$. This limits the number of states we include in the basis,
  as the full matrix ($\Delta H_2 - \Delta H_3$) needs to be inverted over the
  space of tail states. On the contrary, the low-energy diagonalization of
  $H_{l l} + \Delta \protect \mathaccentV {tilde}07E{H}$ is performed
  efficiently via the Lanczos method. This suggests that more efficient
  numerics could be achieved by reducing the number of tails states $\ket {\Psi
  _i}$; see \cite {JSL2} for a discussion.}\BibitemShut {Stop}%
\bibitem [{Note5()}]{Note5}%
  \BibitemOpen
  \bibinfo {note} {In practice we fix ${\protect \cal E}_*$ to the value given
  by the local approximation mentioned below Eq.~(\ref {seriesdh}). Further
  iterative improvements are possible, but their effect is
  negligible.}\BibitemShut {Stop}%
\bibitem [{Note6()}]{Note6}%
  \BibitemOpen
  \bibinfo {note} {By applying the power counting arguments mentioned in
  footonote 21, one can estimate $\Delta \protect \mathaccentV {tilde}07E{H}
  \sim N E_T$, as opposed to $\Delta H_2 \sim N^2 E_T$, therefore taming the
  growth of the matrix elements.}\BibitemShut {Stop}%
\bibitem [{\citenamefont {Coser}\ \emph {et~al.}(2014)\citenamefont {Coser},
  \citenamefont {Beria}, \citenamefont {Brandino}, \citenamefont {Konik},\ and\
  \citenamefont {Mussardo}}]{Coser:2014lla}%
  \BibitemOpen
  \bibfield  {author} {\bibinfo {author} {\bibfnamefont {A.}~\bibnamefont
  {Coser}}, \bibinfo {author} {\bibfnamefont {M.}~\bibnamefont {Beria}},
  \bibinfo {author} {\bibfnamefont {G.~P.}\ \bibnamefont {Brandino}}, \bibinfo
  {author} {\bibfnamefont {R.~M.}\ \bibnamefont {Konik}}, \ and\ \bibinfo
  {author} {\bibfnamefont {G.}~\bibnamefont {Mussardo}},\ }\bibfield  {title}
  {\enquote {\bibinfo {title} {{Truncated Conformal Space Approach for 2D
  Landau-Ginzburg Theories}},}\ }\href {\doibase
  10.1088/1742-5468/2014/12/P12010} {\bibfield  {journal} {\bibinfo  {journal}
  {J. Stat. Mech.}\ }\textbf {\bibinfo {volume} {1412}},\ \bibinfo {pages}
  {P12010} (\bibinfo {year} {2014})},\ \Eprint {http://arxiv.org/abs/1409.1494}
  {arXiv:1409.1494 [hep-th]} \BibitemShut {NoStop}%
\bibitem [{\citenamefont {Bajnok}\ and\ \citenamefont
  {L\'ajer}(2016)}]{Bajnok:2015bgw}%
  \BibitemOpen
  \bibfield  {author} {\bibinfo {author} {\bibfnamefont {Z.}~\bibnamefont
  {Bajnok}}\ and\ \bibinfo {author} {\bibfnamefont {M.}~\bibnamefont
  {L\'ajer}},\ }\bibfield  {title} {\enquote {\bibinfo {title} {{Truncated
  Hilbert space approach to the 2d $\phi^{4}$ theory}},}\ }\href {\doibase
  10.1007/JHEP10(2016)050} {\bibfield  {journal} {\bibinfo  {journal} {JHEP}\
  }\textbf {\bibinfo {volume} {10}},\ \bibinfo {pages} {050} (\bibinfo {year}
  {2016})},\ \Eprint {http://arxiv.org/abs/1512.06901} {arXiv:1512.06901
  [hep-th]} \BibitemShut {NoStop}%
\bibitem [{Note7()}]{Note7}%
  \BibitemOpen
  \bibinfo {note} {To estimate extrapolation errors, we fitted subsamples of
  the full data set, obtained by removing points at low $E_T$ in different
  combinations.}\BibitemShut {Stop}%
\bibitem [{\citenamefont {{L\"uscher}}(1986)}]{Luscher:1985dn}%
  \BibitemOpen
  \bibfield  {author} {\bibinfo {author} {\bibfnamefont {M.}~\bibnamefont
  {{L\"uscher}}},\ }\bibfield  {title} {\enquote {\bibinfo {title} {{Volume
  Dependence of the Energy Spectrum in Massive Quantum Field Theories. 1.
  Stable Particle States}},}\ }\href {\doibase 10.1007/BF01211589} {\bibfield
  {journal} {\bibinfo  {journal} {Commun.Math.Phys.}\ }\textbf {\bibinfo
  {volume} {104}},\ \bibinfo {pages} {177} (\bibinfo {year}
  {1986})}\BibitemShut {NoStop}%
\bibitem [{\citenamefont {Klassen}\ and\ \citenamefont
  {Melzer}(1991{\natexlab{a}})}]{Klassen:1990ub}%
  \BibitemOpen
  \bibfield  {author} {\bibinfo {author} {\bibfnamefont {T.~R.}\ \bibnamefont
  {Klassen}}\ and\ \bibinfo {author} {\bibfnamefont {E.}~\bibnamefont
  {Melzer}},\ }\bibfield  {title} {\enquote {\bibinfo {title} {{On the relation
  between scattering amplitudes and finite size mass corrections in QFT}},}\
  }\href {\doibase 10.1016/0550-3213(91)90566-G} {\bibfield  {journal}
  {\bibinfo  {journal} {Nucl. Phys.}\ }\textbf {\bibinfo {volume} {B362}},\
  \bibinfo {pages} {329--388} (\bibinfo {year}
  {1991}{\natexlab{a}})}\BibitemShut {NoStop}%
\bibitem [{Note8()}]{Note8}%
  \BibitemOpen
  \bibinfo {note} {As a further check of the method, the S-matrix (extracted
  via the volume dependence of the spectrum) could in the future be compared to
  the perturbative prediction for $g \ll 1$. See \cite {Coser:2014lla},
  appendix B.}\BibitemShut {Stop}%
\bibitem [{\citenamefont {Klassen}\ and\ \citenamefont
  {Melzer}(1991{\natexlab{b}})}]{Klassen:1990dx}%
  \BibitemOpen
  \bibfield  {author} {\bibinfo {author} {\bibfnamefont {T.~R.}\ \bibnamefont
  {Klassen}}\ and\ \bibinfo {author} {\bibfnamefont {E.}~\bibnamefont
  {Melzer}},\ }\bibfield  {title} {\enquote {\bibinfo {title} {{The
  Thermodynamics of purely elastic scattering theories and conformal
  perturbation theory}},}\ }\href {\doibase 10.1016/0550-3213(91)90159-U}
  {\bibfield  {journal} {\bibinfo  {journal} {Nucl. Phys.}\ }\textbf {\bibinfo
  {volume} {B350}},\ \bibinfo {pages} {635--689} (\bibinfo {year}
  {1991}{\natexlab{b}})}\BibitemShut {NoStop}%
\bibitem [{\citenamefont {Chang}(1976)}]{Chang:1976ek}%
  \BibitemOpen
  \bibfield  {author} {\bibinfo {author} {\bibfnamefont {S.-J.}\ \bibnamefont
  {Chang}},\ }\bibfield  {title} {\enquote {\bibinfo {title} {{The Existence of
  a Second Order Phase Transition in the Two-Dimensional $\phi^4$ Field
  Theory}},}\ }\href {\doibase 10.1103/PhysRevD.13.2778,
  10.1103/PhysRevD.16.1979} {\bibfield  {journal} {\bibinfo  {journal}
  {Phys.Rev.}\ }\textbf {\bibinfo {volume} {D13}},\ \bibinfo {pages} {2778}
  (\bibinfo {year} {1976})}\BibitemShut {NoStop}%
\bibitem [{\citenamefont {Schaich}\ and\ \citenamefont
  {Loinaz}(2009)}]{Schaich:2009jk}%
  \BibitemOpen
  \bibfield  {author} {\bibinfo {author} {\bibfnamefont {D.}~\bibnamefont
  {Schaich}}\ and\ \bibinfo {author} {\bibfnamefont {W.}~\bibnamefont
  {Loinaz}},\ }\bibfield  {title} {\enquote {\bibinfo {title} {{An improved
  lattice measurement of the critical coupling in $\phi_2^4$ theory}},}\ }\href
  {\doibase 10.1103/PhysRevD.79.056008} {\bibfield  {journal} {\bibinfo
  {journal} {Phys.Rev.}\ }\textbf {\bibinfo {volume} {D79}},\ \bibinfo {pages}
  {056008} (\bibinfo {year} {2009})},\ \Eprint {http://arxiv.org/abs/0902.0045}
  {arXiv:0902.0045 [hep-lat]} \BibitemShut {NoStop}%
\bibitem [{\citenamefont {Milsted}\ \emph {et~al.}(2013)\citenamefont
  {Milsted}, \citenamefont {Haegeman},\ and\ \citenamefont
  {Osborne}}]{Milsted:2013rxa}%
  \BibitemOpen
  \bibfield  {author} {\bibinfo {author} {\bibfnamefont {A.}~\bibnamefont
  {Milsted}}, \bibinfo {author} {\bibfnamefont {J.}~\bibnamefont {Haegeman}}, \
  and\ \bibinfo {author} {\bibfnamefont {T.~J.}\ \bibnamefont {Osborne}},\
  }\bibfield  {title} {\enquote {\bibinfo {title} {{Matrix product states and
  variational methods applied to critical quantum field theory}},}\ }\href
  {\doibase 10.1103/PhysRevD.88.085030} {\bibfield  {journal} {\bibinfo
  {journal} {Phys.Rev.}\ }\textbf {\bibinfo {volume} {D88}},\ \bibinfo {pages}
  {085030} (\bibinfo {year} {2013})},\ \Eprint {http://arxiv.org/abs/1302.5582}
  {arXiv:1302.5582 [hep-lat]} \BibitemShut {NoStop}%
\bibitem [{\citenamefont {Bosetti}\ \emph {et~al.}(2015)\citenamefont
  {Bosetti}, \citenamefont {De~Palma},\ and\ \citenamefont
  {Guagnelli}}]{Bosetti:2015lsa}%
  \BibitemOpen
  \bibfield  {author} {\bibinfo {author} {\bibfnamefont {P.}~\bibnamefont
  {Bosetti}}, \bibinfo {author} {\bibfnamefont {B.}~\bibnamefont {De~Palma}}, \
  and\ \bibinfo {author} {\bibfnamefont {M.}~\bibnamefont {Guagnelli}},\
  }\bibfield  {title} {\enquote {\bibinfo {title} {{Monte Carlo determination
  of the critical coupling in $\phi^4_2$ theory}},}\ }\href {\doibase
  10.1103/PhysRevD.92.034509} {\bibfield  {journal} {\bibinfo  {journal} {Phys.
  Rev.}\ }\textbf {\bibinfo {volume} {D92}},\ \bibinfo {pages} {034509}
  (\bibinfo {year} {2015})},\ \Eprint {http://arxiv.org/abs/1506.08587}
  {arXiv:1506.08587} \BibitemShut {NoStop}%
\bibitem [{\citenamefont {Pelissetto}\ and\ \citenamefont
  {Vicari}(2015)}]{Pelissetto:2015yha}%
  \BibitemOpen
  \bibfield  {author} {\bibinfo {author} {\bibfnamefont {A.}~\bibnamefont
  {Pelissetto}}\ and\ \bibinfo {author} {\bibfnamefont {E.}~\bibnamefont
  {Vicari}},\ }\bibfield  {title} {\enquote {\bibinfo {title} {{Critical mass
  renormalization in renormalized $\phi^4$ theories in two and three
  dimensions}},}\ }\href {\doibase 10.1016/j.physletb.2015.11.015} {\bibfield
  {journal} {\bibinfo  {journal} {Phys. Lett.}\ }\textbf {\bibinfo {volume}
  {B751}},\ \bibinfo {pages} {532--534} (\bibinfo {year} {2015})},\ \Eprint
  {http://arxiv.org/abs/1508.00989} {arXiv:1508.00989 [hep-th]} \BibitemShut
  {NoStop}%
\bibitem [{Note9()}]{Note9}%
  \BibitemOpen
  \bibinfo {note} {To compare with the critical coupling extractions using the
  light front quantization \cite
  {Chabysheva:2015ynr,Burkardt:2016ffk,Anand:2017yij} one has to perform
  nonperturbative mass renormalization \cite {Burkardt:2016ffk}.}\BibitemShut
  {Stop}%
\bibitem [{Note10()}]{Note10}%
  \BibitemOpen
  \bibinfo {note} {The central value corresponds to the smallest $\chi ^2(g_c)=
  \DOTSB \sum@ \slimits@ _{i=1}^N {(y_i-f(x_i))^2}/{err_i}^2$. The uncertainty
  interval was conservatively determined from the condition $\protect \sqrt
  {\chi ^2(g_c)}\leqslant 3\protect \sqrt {\chi ^2(2.76)}$. Our determination
  is the best HT measurement of $g_c$. It is compatible with and has accuracy
  comparable to other available determinations \cite {JSL2}.}\BibitemShut
  {Stop}%
\bibitem [{\citenamefont {Brodsky}\ \emph {et~al.}(1998)\citenamefont
  {Brodsky}, \citenamefont {Pauli},\ and\ \citenamefont
  {Pinsky}}]{Brodsky:1997de}%
  \BibitemOpen
  \bibfield  {author} {\bibinfo {author} {\bibfnamefont {S.~J.}\ \bibnamefont
  {Brodsky}}, \bibinfo {author} {\bibfnamefont {H.-C.}\ \bibnamefont {Pauli}},
  \ and\ \bibinfo {author} {\bibfnamefont {S.~S.}\ \bibnamefont {Pinsky}},\
  }\bibfield  {title} {\enquote {\bibinfo {title} {{Quantum chromodynamics and
  other field theories on the light cone}},}\ }\href {\doibase
  10.1016/S0370-1573(97)00089-6} {\bibfield  {journal} {\bibinfo  {journal}
  {Phys.Rept.}\ }\textbf {\bibinfo {volume} {301}},\ \bibinfo {pages}
  {299--486} (\bibinfo {year} {1998})},\ \Eprint
  {http://arxiv.org/abs/hep-ph/9705477} {arXiv:hep-ph/9705477 [hep-ph]}
  \BibitemShut {NoStop}%
\bibitem [{\citenamefont {Katz}\ \emph
  {et~al.}(2014{\natexlab{a}})\citenamefont {Katz}, \citenamefont {Tavares},\
  and\ \citenamefont {Xu}}]{Katz:2013qua}%
  \BibitemOpen
  \bibfield  {author} {\bibinfo {author} {\bibfnamefont {E.}~\bibnamefont
  {Katz}}, \bibinfo {author} {\bibfnamefont {G.~M.}\ \bibnamefont {Tavares}}, \
  and\ \bibinfo {author} {\bibfnamefont {Y.}~\bibnamefont {Xu}},\ }\bibfield
  {title} {\enquote {\bibinfo {title} {{Solving 2D QCD with an adjoint fermion
  analytically}},}\ }\href {\doibase 10.1007/JHEP05(2014)143} {\bibfield
  {journal} {\bibinfo  {journal} {JHEP}\ }\textbf {\bibinfo {volume} {1405}},\
  \bibinfo {pages} {143} (\bibinfo {year} {2014}{\natexlab{a}})},\ \Eprint
  {http://arxiv.org/abs/1308.4980} {arXiv:1308.4980 [hep-th]} \BibitemShut
  {NoStop}%
\bibitem [{\citenamefont {Katz}\ \emph
  {et~al.}(2014{\natexlab{b}})\citenamefont {Katz}, \citenamefont {Tavares},\
  and\ \citenamefont {Xu}}]{Katz:2014uoa}%
  \BibitemOpen
  \bibfield  {author} {\bibinfo {author} {\bibfnamefont {E.}~\bibnamefont
  {Katz}}, \bibinfo {author} {\bibfnamefont {G.~M.}\ \bibnamefont {Tavares}}, \
  and\ \bibinfo {author} {\bibfnamefont {Y.}~\bibnamefont {Xu}},\ }\bibfield
  {title} {\enquote {\bibinfo {title} {{A solution of 2D QCD at Finite $N$
  using a conformal basis}},}\ }\href@noop {} {\  (\bibinfo {year}
  {2014}{\natexlab{b}})},\ \Eprint {http://arxiv.org/abs/1405.6727}
  {arXiv:1405.6727 [hep-th]} \BibitemShut {NoStop}%
\bibitem [{\citenamefont {Chabysheva}(2016)}]{Chabysheva:2015ynr}%
  \BibitemOpen
  \bibfield  {author} {\bibinfo {author} {\bibfnamefont {S.~S.}\ \bibnamefont
  {Chabysheva}},\ }\bibfield  {title} {\enquote {\bibinfo {title} {{Light-front
  $\phi^4_{1+1}$ theory using a many-boson symmetric-polynomial basis}},}\
  }\bibfield  {booktitle} {\emph {\bibinfo {booktitle} {{Proceedings, Light
  Cone 2015: Frascati, Italy, Sep 21-25, 2015}}},\ }\href {\doibase
  10.1007/s00601-016-1106-0} {\bibfield  {journal} {\bibinfo  {journal} {Few
  Body Syst.}\ }\textbf {\bibinfo {volume} {57}},\ \bibinfo {pages} {675--680}
  (\bibinfo {year} {2016})},\ \Eprint {http://arxiv.org/abs/1512.08770}
  {arXiv:1512.08770 [hep-ph]} \BibitemShut {NoStop}%
\bibitem [{\citenamefont {Burkardt}\ \emph {et~al.}(2016)\citenamefont
  {Burkardt}, \citenamefont {Chabysheva},\ and\ \citenamefont
  {Hiller}}]{Burkardt:2016ffk}%
  \BibitemOpen
  \bibfield  {author} {\bibinfo {author} {\bibfnamefont {M.}~\bibnamefont
  {Burkardt}}, \bibinfo {author} {\bibfnamefont {S.~S.}\ \bibnamefont
  {Chabysheva}}, \ and\ \bibinfo {author} {\bibfnamefont {J.~R.}\ \bibnamefont
  {Hiller}},\ }\bibfield  {title} {\enquote {\bibinfo {title} {{Two-dimensional
  light-front $\phi^4$ theory in a symmetric polynomial basis}},}\ }\href@noop
  {} {\  (\bibinfo {year} {2016})},\ \Eprint {http://arxiv.org/abs/1607.00026}
  {arXiv:1607.00026 [hep-th]} \BibitemShut {NoStop}%
\bibitem [{\citenamefont {Katz}\ \emph {et~al.}(2016)\citenamefont {Katz},
  \citenamefont {Khandker},\ and\ \citenamefont {Walters}}]{Katz:2016hxp}%
  \BibitemOpen
  \bibfield  {author} {\bibinfo {author} {\bibfnamefont {E.}~\bibnamefont
  {Katz}}, \bibinfo {author} {\bibfnamefont {Z.~U.}\ \bibnamefont {Khandker}},
  \ and\ \bibinfo {author} {\bibfnamefont {M.~T.}\ \bibnamefont {Walters}},\
  }\bibfield  {title} {\enquote {\bibinfo {title} {{A Conformal Truncation
  Framework for Infinite-Volume Dynamics}},}\ }\href {\doibase
  10.1007/JHEP07(2016)140} {\bibfield  {journal} {\bibinfo  {journal} {JHEP}\
  }\textbf {\bibinfo {volume} {07}},\ \bibinfo {pages} {140} (\bibinfo {year}
  {2016})},\ \Eprint {http://arxiv.org/abs/1604.01766} {arXiv:1604.01766
  [hep-th]} \BibitemShut {NoStop}%
\bibitem [{\citenamefont {Anand}\ \emph {et~al.}(2017)\citenamefont {Anand},
  \citenamefont {Genest}, \citenamefont {Katz}, \citenamefont {Khandker},\ and\
  \citenamefont {Walters}}]{Anand:2017yij}%
  \BibitemOpen
  \bibfield  {author} {\bibinfo {author} {\bibfnamefont {N.}~\bibnamefont
  {Anand}}, \bibinfo {author} {\bibfnamefont {V.~X.}\ \bibnamefont {Genest}},
  \bibinfo {author} {\bibfnamefont {E.}~\bibnamefont {Katz}}, \bibinfo {author}
  {\bibfnamefont {Z.~U.}\ \bibnamefont {Khandker}}, \ and\ \bibinfo {author}
  {\bibfnamefont {M.~T.}\ \bibnamefont {Walters}},\ }\bibfield  {title}
  {\enquote {\bibinfo {title} {{RG Flow from $\phi^4$ Theory to the 2D Ising
  Model}},}\ }\href@noop {} {\  (\bibinfo {year} {2017})},\ \Eprint
  {http://arxiv.org/abs/1704.04500} {arXiv:1704.04500 [hep-th]} \BibitemShut
  {NoStop}%
\bibitem [{\citenamefont {Konik}\ and\ \citenamefont
  {Adamov}(2007)}]{Konik:2007cb}%
  \BibitemOpen
  \bibfield  {author} {\bibinfo {author} {\bibfnamefont {R.~M.}\ \bibnamefont
  {Konik}}\ and\ \bibinfo {author} {\bibfnamefont {Y.}~\bibnamefont {Adamov}},\
  }\bibfield  {title} {\enquote {\bibinfo {title} {Numerical renormalization
  group for continuum one-dimensional systems},}\ }\href {\doibase
  10.1103/PhysRevLett.98.147205} {\bibfield  {journal} {\bibinfo  {journal}
  {Phys. Rev. Lett.}\ }\textbf {\bibinfo {volume} {98}},\ \bibinfo {pages}
  {147205} (\bibinfo {year} {2007})},\ \Eprint
  {http://arxiv.org/abs/cond-mat/0701605} {arXiv:cond-mat/0701605
  [cond-mat.str-el]} \BibitemShut {NoStop}%
\bibitem [{\citenamefont {Wilson}(1975)}]{Wilson}%
  \BibitemOpen
  \bibfield  {author} {\bibinfo {author} {\bibfnamefont {K.~G.}\ \bibnamefont
  {Wilson}},\ }\bibfield  {title} {\enquote {\bibinfo {title} {The
  renormalization group: Critical phenomena and the kondo problem},}\ }\href
  {\doibase 10.1103/RevModPhys.47.773} {\bibfield  {journal} {\bibinfo
  {journal} {Rev. Mod. Phys.}\ }\textbf {\bibinfo {volume} {47}},\ \bibinfo
  {pages} {773--840} (\bibinfo {year} {1975})}\BibitemShut {NoStop}%
\end{thebibliography}%

\end{document}